\documentclass[10pt,a4paper]{article}
\usepackage[nottoc]{tocbibind}
\usepackage[toc,page]{appendix}
\usepackage[dvipsnames]{xcolor}
\usepackage{physics}
\usepackage{tablefootnote}
\usepackage{caption}
\usepackage{subcaption}
\usepackage{adjustbox}
\usepackage{mathtools}

\usepackage[utf8]{inputenc}   
\usepackage[T1]{fontenc}      
\usepackage{graphicx}         
\usepackage{amssymb}          
\usepackage{amsmath}          
\usepackage{slashed}          
\usepackage{a4wide}           
\usepackage[font=small]{caption}   
\usepackage{url}              
\usepackage{booktabs}         
\usepackage[output-decimal-marker={,},list-final-separator={ en }]{siunitx}          

\usepackage[style=ieee,citestyle=numeric-comp,backend=bibtex]{biblatex}
\usepackage[colorlinks=true,linkcolor=black,urlcolor=blue,citecolor=blue]{hyperref}

\usepackage{epigraph}

\usepackage[dvipsnames]{xcolor}

\numberwithin{equation}{section}

\usepackage{authblk}

\title{Probing the QCD $\bar \theta$ term with paramagnetic molecules }
\author[$a$,$b$]{Heleen Mulder}
\author[$a$,$b$]{Rob Timmermans}
\author[$a$,$c$]{Jordy de Vries}
\affil[$a$]{Nikhef, National Institute for Subatomic Physics, The Netherlands}
\affil[$b$]{Van Swinderen Institute for Particle Physics and Gravity, University of Groningen,
The Netherlands}
\affil[$c$]{
Institute of Physics and Delta Institute for Theoretical Physics, University of Amsterdam, The Netherlands}

\begin{document}

\maketitle

\begin{abstract}

    The experimental search for CP violation in paramagnetic atomic and molecular systems has made impressive progress in recent years. This has led to strong upper limits on the electron electric dipole moment. The same measurements can also be used to constrain hadronic sources of CP violation through CP-violating interactions between the electrons and the nucleus. We employ heavy-baryon chiral perturbation theory to compute such CP-violating semileptonic electron-nucleus interactions arising from the QCD theta term. We sharpen earlier results by determining the relevant short-distance effects and by an explicit two-loop calculation of meson-photon diagrams. We derive a bound of $|\bar{\theta}| < 1.5 \cdot 10^{-8}$ at $90\%$ confidence, based on HfF$^+$ experiments at JILA. A further experimental improvement of one to two orders of magnitude would make paramagnetic molecular electric dipole moment experiments competitive with the neutron and diamagnetic atom program in constraining strong CP violation and higher-dimensional CP-odd quark-gluon operators.
\end{abstract}

\section{Introduction}\label{sec:intro}
Cosmological observations indicate an order $10^{10}$ asymmetry between the amount of baryons and antibaryons in our universe. As pointed out by Sakharov \cite{Sakharov:1967dj}, a dynamical resolution of this asymmetry requires the violation of charge conjugation and parity (CP) or, equivalently, the violation of time-reversal symmetry\footnote{Throughout this paper, we assume that CPT is conserved. Hence, $\slashed{P}$ and $\slashed{T}$ sources are sources of CP violation.}. The Standard Model (SM) of particle physics without neutrino masses contains two sources of CP violation. The first is the Cabibbo-Kobayashi-Maskawa (CKM) phase, which leads to CP violation in flavor-changing processes and has been measured extensively in the $B$ and $K$ meson sectors. The second is the QCD vacuum angle $\bar{\theta}$, the value of which has not yet been measured, only constrained by upper bounds. Currently, null measurements of the neutron electric dipole moment (EDM) provide the tightest bound, namely $|\bar{\theta}| < 10^{-10}$ \cite{Abel:2020pzs,Pospelov:1999mv,Liang:2023jfj}. The question of why this parameter is so small constitutes the strong CP problem \cite{PospelovRitz2005,Di_Luzio_2020}.

Beyond-the-Standard-Model (BSM) sources of CP violation are necessary to explain a dynamical origin of the observed matter-antimatter asymmetry \cite{Morrissey:2012db}. EDMs provide excellent probes to look for these yet unknown sources \cite{PospelovRitz2005,Engel:2013lsa}. In particular, EDM searches provide a window on BSM and $\bar{\theta}$ flavor-diagonal CP violation which can be regarded as "background-free", due to the negligible contribution of the CKM phase (although in the case of paramagnetic EDMs, the CKM contribution was found to be larger than previously believed \cite{Ema:2022yra}). In addition, the increasingly precise null results obtained in a variety of EDM searches set stringent limits on general BSM CP violation not necessarily linked to the matter-antimatter asymmetry, see e.g.~Refs.~\cite{Cirigliano:2016nyn, deVries:2017ncy,Kley:2021yhn,Kumar:2024yuu} for effective field theory (EFT) studies.

The atomic and molecular systems on which EDM experiments are performed are usually divided into paramagnetic and diamagnetic species, which respectively have non-zero and zero total electron spin. In principle, paramagnetic systems are more sensitive to leptonic CP violation, through the spin of the unpaired electron, while diamagnetic systems are more suitable as probes of hadronic CP violation. However, the experimental precision of the searches using paramagnetic molecules has improved much faster than that of diamagnetic and neutron EDM searches. Over the last two decades, the upper bound on the electron EDM $d_e$ $-$ determined using the heavy paramagnetic molecules YbF, ThO and HfF$^+$ $-$ has improved by a factor of about 400 \cite{Hudson:2011zz,ACME:2013pal,Cairncross:2017fip,ACME:2018yjb,Roussy:2022cmp} compared to the best atomic limit from ${}^{205}$Tl \cite{Regan:2002ta}, while the upper bound on the neutron EDM improved by a factor of about 4 in the same period \cite{Harris:1999jx,Baker:2006ts,Serebrov:2013tba,Abel:2020pzs}. Similarly, the EDM limits on diamagnetic atoms such as ${}^{126}$Xe \cite{PhysRevLett.86.22,Sachdeva:2019rkt,Allmendinger:2019jrk}, ${}^{199}$Hg \cite{Griffith:2009zz,Graner:2016ses}, and ${}^{225}$Ra \cite{Parker:2015yka,Bishof:2016uqx} are only slowly improving. Moreover, experiments with paramagnetic molecules are expected to further improve by orders of magnitude \cite{ACME:2018yjb,Roussy:2022cmp,NL-eEDM:2018lno,Vutha:2018tsz,Ho:2023xuo}.

Considering the spectacular progress and the prospects for EDM searches with paramagnetic molecules, it is becoming increasingly relevant to investigate whether hadronic CP violation can be probed through paramagnetic EDMs. In particular, hadronic CP violation can lead to semileptonic CP-odd operators such as
\begin{equation}
    \mathcal{L} = C_{\mathrm{SP}}^p \frac{G_F}{\sqrt{2}}\bar{e}i\gamma_5 e\,\bar{p}p + C_{\mathrm{SP}}^n \frac{G_F}{\sqrt{2}}\bar{e}i\gamma_5 e\,\bar{n}n\,,
    \label{eq:CSPLagrangian}
\end{equation}
 where SP refers to the scalar nucleon bilinear and the pseudoscalar electron bilinear. In Eq.~\eqref{eq:CSPLagrangian}, the nucleon fields $p$ and $n$ are non-relativistic fields in the framework of heavy-baryon chiral perturbation theory (see Sec.~\ref{sec:chiPT}), while the electrons are described by relativistic fields. The operators $C_{\mathrm{SP}}^N$ couple to the electron spin, and, being independent of the nucleon spin, their effect scales with the number of nucleons. These characteristics lead to the expectation that the $C_{\mathrm{SP}}$ coupling is a promising route for hadronic CP violation to contribute to EDMs of (heavy) paramagnetic molecules.

In this work, we compute the values of $C^{p,n}_\text{SP}$ induced by CP-violating meson-nucleon interactions. 
For CP-violating operators involving quarks and gluons, such as the QCD theta term, CP-odd meson-nucleon interactions arise at leading order (LO) in the framework of chiral perturbation theory ($\chi$PT) \cite{Mereghetti:2010tp,deVries:2012ab,Bsaisou:2014oka}. For the purpose of this work, the most important interactions are 
\begin{equation}
    \mathcal{L}_{\pi / \eta N\!N} = \bar{g}_0 \bar{N} \tau^a N \pi^a + \bar{g}_1 \bar{N}\!N \pi^0 + \bar{g}_{0\eta} \bar{N}\!N \eta\,,
    \label{eq:gbarterms}
\end{equation}
in terms of the nucleon doublet $N=(p\,n)^T$, the pion triplet $\pi^a$, and the $\eta$ meson. The coupling constants $\bar g_{0,1,0\eta}$, often called low-energy constants (LECs) in the $\chi$PT literature, are functions of the underlying quark-gluon source of CP violation. We focus on the QCD $\bar \theta$ term, where the LECs are relatively well known as functions of $\bar \theta$ \cite{Crewther:1979pi,deVries:2015una}, but our results can readily be applied to other CP-odd mechanisms once values of the $\bar g_{0,1,0\eta}$ are known. 

In Ref.~\cite{PospelovThO} it was shown that the semileptonic interactions $C^{p,n}_\text{SP}$ arise from one- and two-loop meson-photon diagrams involving one vertex from Eq.~\eqref{eq:gbarterms}. While this leads to sizeable contributions to paramagnetic EDMs, the calculation also involves several sources of theoretical uncertainty. For instance, both the LO and the next-to-leading-order (NLO) diagrams as computed in Ref.~\cite{PospelovThO} diverge and depend on the minimal-subtraction renormalization scale $\mu$. Second, there exists an accidental cancellation between formally LO and NLO diagrams, making it important to identify all NLO corrections. Our goal here is therefore to improve the theoretical underpinning of the important results of Ref.~\cite{PospelovThO}, in a model-independent and systematic EFT framework. 

Furthermore, obtaining a robust upper bound on $\bar{\theta}$ from paramagnetic EDM searches, independent of the neutron EDM bound, would be very valuable. The calculation of the neutron EDM in terms of the $\bar \theta$ term has been notoriously difficult. Chiral methods are not reliable, as the calculable long-distance pionic contributions must be renormalized by unknown short-distance counterterms \cite{Crewther:1979pi, Hockings:2005cn}. Only in recent years have lattice QCD calculations provided the first non-zero results, but the uncertainties are still uncomfortably large \cite{Dragos:2019oxn,Liang:2023jfj,Liu:2024kqy} (see also Ref.~\cite{Ema:2024vfn} for other sources of discomfort). A potentially cleaner and hopefully in the future competitive constraint from paramagnetic systems would be welcome.

This paper is organized as follows. We introduce the chiral Lagrangian in Sec.~\ref{sec:chiPT}, and discuss the values of the CP-odd couplings in case of the QCD $\bar \theta$ term. In Sec.~\ref{sec:calculations}, we explicitly compute the one- and two-loop contributions and demonstrate that our results are properly renormalized. We then combine all contributions in Sec.~\ref{sec:combinedsummary}, providing expressions for $C_\text{SP}$ in terms of hadronic CP-odd couplings as well as the specific hadronic CP-violating source $\bar{\theta}$. We also deduce an upper bound on $\bar \theta$. Sec.~\ref{sec:conclusions} is reserved for our conclusions and directions for future work.

\section{The chiral Lagrangian and power counting}\label{sec:chiPT}

To compute the contributions to $C_{\text{SP}}^{n,p}$ from the meson-exchange (ME) and NLO pion-loop (PL) diagrams in, respectively, Fig.~\ref{fig:LOdiagramforpaper} and \ref{fig:fullNLOdiagrams}, we will apply chiral perturbation theory ($\chi$PT), the low-energy EFT of QCD \cite{Weinberg:1978kz,Gasser:1983yg}. The $\chi$PT Lagrangian contains all terms involving the relevant low-energy degrees of freedom consistent with QCD supplemented, in our case, by CP-violating operators. The big advantage of $\chi$PT is that low-energy observables can be computed in an expansion in $Q/\Lambda_\chi$, where $Q$ is the typical momentum scale of the observable and $\Lambda_\chi \sim 1$ GeV, the chiral-breaking scale. 
$\chi$PT can be extended to include nucleons, but since $m_N \sim \Lambda_
\chi$, the power counting is more complicated. We consider non-relativistic nucleons and apply heavy-baryon $\chi$PT (HB$\chi$PT) \cite{Jenkins:1990jv} to overcome this difficulty. 

The LO HB$\chi$PT Lagrangian density is given by
\begin{equation}
    \mathcal{L}^{(0)} = \frac{1}{2} D_\mu \boldsymbol{\pi} \cdot D^\mu \boldsymbol{\pi} - \frac{m_\pi^2}{2} \boldsymbol{\pi}^2+\bar{N}\left( iv \cdot \mathcal{D} - \frac{2 g_A}{F_\pi} S^\mu \boldsymbol{\tau} \cdot D_\mu \boldsymbol{\pi} \right)N + ...\,,
    \label{eq:LOchiPTLagrangian}
\end{equation}
 where $F_\pi \simeq 92.1$ MeV is the pion decay constant \cite{FlavourLatticeAveragingGroupFLAG:2024oxs}, $v^\mu$ and $S^\mu$ denote the nucleon velocity and spin, respectively, and the dots denote interactions involving more pion fields. We only require the electromagnetic part of the covariant derivatives,
\begin{align}
\begin{split}
    (D_\mu \pi)_i &= (\partial_\mu \delta_{ij} + e A_\mu \epsilon_{3ij}) \pi_j + ...\,, \\
    \mathcal{D}_\mu N &= \left( \partial_\mu + \frac{ie}{2} A_\mu (1+\tau_3)\right)N + ...\,,
    \label{eq:DmuEMchiPT}
\end{split}
\end{align}
where again the dots denote terms with more pion fields, which we do not need in this work. Additionally, $g_A \simeq 1.27$ is the (CP-even) pion-nucleon axial-vector coupling \cite{PDG}, $m_\pi$ is the pion mass, and $e > 0$ is the proton charge. For the PL diagrams in Fig.~\ref{fig:fullNLOdiagrams}, we also need the nucleon anomalous magnetic moment vertex, which appears at next-to-leading order in the HB$\chi$PT Lagrangian density \cite{Bernard:1995dp}
\begin{equation}
    \mathcal{L}^{(1)} \supset \frac{e}{4 m_N} \varepsilon^{\alpha \beta \mu \nu} v_\alpha \bar{N} S_\beta [(1+\kappa_0)+(1+\kappa_1)\tau_3] N F_{\mu \nu}\,,
    \label{eq:anommagnmom}
\end{equation}
where $\kappa_0 =-0.12$ and $\kappa_1=3.7$ are the isoscalar and isovector nucleon anomalous magnetic moment. $\kappa_{0,1}$ are related to the nucleon magnetic moments through $\mu_p / \mu_N = 1 + \frac{1}{2}(\kappa_0 + \kappa_1) \simeq 2.79$ and $\mu_n / \mu_N = \frac{1}{2}(\kappa_0 - \kappa_1)\simeq -1.91$ in units of the nuclear magneton $\mu_N$.

 In the ME calculation, we use the Wess-Zumino (WZ) Lagrangian term for the anomalous meson-photon-photon couplings \cite{Savage:1992ac},
\begin{equation}
    \mathcal{L}_\text{WZ} = \frac{\alpha}{8 \pi F_\pi} \varepsilon_{\mu \nu \lambda \sigma} F^{\mu \nu} F^{\lambda \sigma} \left( \pi^0 + \frac{F_\pi}{F_\eta}\frac{\eta}{\sqrt{3}} \right)\,,
    \label{eq:WZterm}
\end{equation}
where $\epsilon^{0123} = +1$, $\eta_{\mu\nu} = (+,-,-,-)$ and $F_\eta \simeq 1.3 F_\pi$ \cite{Gasser:1984gg,Borasoy:2003yb}. We also require
\begin{equation}
    \mathcal{L}_{\text{CT}} = -\frac{ \alpha^2 m_l}{2 \pi^2 F_\pi}\, \bar{l} i\gamma^5 l \left(\chi^\pi \pi^0 + \chi^\eta \frac{F_\pi}{F_\eta} \frac{\eta}{\sqrt{3}} \right)
    \,,\label{eq:CT}
\end{equation}
where $l =\{e,\,\mu\}$. 
Eq.~\eqref{eq:CT} describes a local coupling of the $\pi^0 / \eta$ to the electron line (the grey dot in Fig.~\ref{fig:LOdiagramforpaperCT}), which regularizes the UV-divergent two-photon loop in the ME case (Fig.~\ref{fig:LOdiagramforpaperloop}) \cite{Savage:1992ac}. By Naive Dimensional Analysis (NDA) we estimate $\chi^\pi \sim\chi^\eta= \mathcal O(1)$ \cite{Manohar:1983md}. Furthermore, $SU(3)$ $\chi$PT predicts $\chi^\pi = \chi^\eta$. However, the values of $\chi$ can be determined by fitting to the observed $\pi^0 \rightarrow e^+ e^-$ and $\eta \rightarrow \mu^+ \mu^-$ branching ratios, see Sec.~\ref{subsec:LO}, so we will treat $\chi^\pi$ and $\chi^\eta$ separately.

Based on the above Lagrangian, we now power count the order of the diagrams in Fig.~\ref{fig:LOdiagramforpaperloop} and \ref{fig:fullNLOdiagrams}. It is relevant to note that these diagrams, and in fact all diagrams that arise from hadronic CP violation and (potentially) contribute to $C_\text{SP}$, involve the exchange of two photons between the electron and nucleon lines. This is necessary in order to obtain the scalar-pseudoscalar structure shown in Eq.~\eqref{eq:CSPLagrangian}. One-photon-exchange diagrams would result in CP-violating electron-nucleon interactions relevant for diamagnetic systems (for instance through the nuclear EDM or Schiff moment) where CP violation is associated with the nucleus instead of the electron. 

In diagram \ref{fig:LOdiagramforpaperloop}, the loop momentum is ultrasoft $Q_{\mathrm{us}}\sim m_e \ll m_{\pi,\eta}$, and the integral brings in a factor of $Q^4_{\mathrm{us}}/(4\pi)^2$. The internal photon and electron propagators scale, respectively, as  $1/Q_{\mathrm{us}}^2$ and $m_e/Q_{\mathrm{us}}^2$, where we keep the electron mass to account for the chirality flip required by the interactions in Eq.~\eqref{eq:CSPLagrangian}. The meson propagator reduces to $1/m_{\pi,\eta}^2$. Putting this together gives
\begin{eqnarray}\label{PCLO}
G_F\,C_{\mathrm{SP}}^{n,p}(\mathrm{ME})\sim \frac{m_e \alpha^2}{(4\pi)^2 F_\pi}\left\{\frac{\bar g_0}{m_\pi^2},\,\frac{\bar g_1}{m_\pi^2},\,\frac{\bar g_{0\eta}}{m_\eta^2}\right\}\,,
\end{eqnarray}
depending on which CP-odd meson-nucleon coupling appears in diagram \ref{fig:LOdiagramforpaperloop}. There is no loop integral in diagram \ref{fig:LOdiagramforpaperCT}, so the contribution can be read directly from the vertices.

If we specify the underlying CP-violating mechanism at the quark-gluon level, it becomes possible to further separate the various terms in Eq.~\eqref{PCLO} \cite{deVries:2012ab}. For the QCD $\bar \theta$ term, the ratio $\bar g_1/\bar g_0$ is suppressed because the $\bar \theta$ term conserves isospin symmetry. While the CP-odd $\eta$-nucleon coupling conserves isospin, the large strange quark mass implies that the contribution from $\bar g_{0\eta}$ is suppressed with respect to $\bar g_0$. Below we will make more quantitative statements.

We next consider the two-loop diagrams in Fig.~\ref{fig:fullNLOdiagrams}. In both loops, the loop momentum is soft $Q\sim m_\pi \sim F_\pi$, and each integral brings in a factor $Q^4/(4\pi)^2$. Photon propagators now scale as $1/Q^2$, and the electron propagator as $m_e/Q^2$ because of the required chirality flip. The meson propagators scale as $1/Q^2$ and the heavy-nucleon propagators as $1/Q$. As discussed in more detail below, the photon-nucleon vertex must be magnetic and thus scales as $e Q/m_N$. All together, we obtain
\begin{eqnarray}\label{PCNLO}
G_F\,C_{\mathrm{SP}}^{n,p}(\mathrm{PL})\sim \frac{m_e \alpha^2}{(4\pi)^2 F_\pi}\left(\frac{g_A \bar g_0}{Q m_N}\right)\,,
\end{eqnarray}
where now only $\bar g_0$ enters because we require a charged meson in the loop. Comparing Eqs.~\eqref{PCLO} and \eqref{PCNLO}, using $g_A \sim 1$ and $Q\sim m_\pi$, confirms that the PL diagrams are suppressed by one power of $m_\pi/m_N$. 

Apart from meson loops, we can write down direct short-distance contributions to $C_\mathrm{SP}$. The relative size of such counterterms with respect to the pion diagrams above depends on the underlying source of CP violation. For example for the $\bar \theta$ term, NDA would assign a scaling
\begin{equation}
G_F\, C_{\mathrm{SP},\mathrm{CT}} = \mathcal O\left(\frac{m_e \alpha^2}{\Lambda_\chi^4}m_*\bar \theta\right)\,,
\end{equation}
where $\Lambda_\chi \sim m_N \sim 4\pi F_\pi$ is the $\chi$PT breakdown scale. Using $\bar g_0 =\mathcal O(m_* \bar \theta/F_\pi)$ in Eq.~\eqref{PCNLO} gives
\begin{equation}
G_F\,C_{\mathrm{SP}}(\mathrm{PL})=\mathcal O\left(\frac{m_e \alpha^2}{\Lambda_\chi^3 Q }m_*\bar \theta\right)\,,
\end{equation}
and thus the counterterm is down by $Q/\Lambda_\chi$ and beyond the accuracy of our calculation. 

Before proceeding to the calculations of the ME and PL contributions, we note that another potentially significant contribution to $C_\mathrm{SP}$ comes from two-photon-exchange box diagrams involving the nucleon EDM and the nucleon magnetic dipole moment. This diagram only involves ultrasoft energy scales and no virtual mesons. It is sensitive to nuclear excited states. Ref.~\cite{PospelovThO} estimated the contributions from this diagram using a naive Fermi-gas model for the nucleus, finding a comparable contribution as from the PL diagrams. A slightly larger result was found in Ref.~\cite{Flambaum:2020gou}. The uncertainty of these calculations are hard to assess and we therefore leave an EFT approach of the box diagrams to future work \cite{BoxFuture}.

\subsection{Values of CP-odd LECs for the QCD \texorpdfstring{$\bar \theta$}{Theta} term}\label{sec:thetaLECs}

Our first goal will be to compute values of $C_{\mathrm{SP}}^{n,p}$ in terms of the LECs appearing in Eq.~\eqref{eq:gbarterms}. However, to connect to the underlying source of CP violation we need to express these LECs in terms of CP-violating operators at the level of elementary fields. This is a notoriously difficult problem, and at present for most CP-odd operators $-$ such as the quark chromo-electric dipole moment, CP-odd four-quark operators, or the Weinberg three-gluon operator $-$ the values of the LECs suffer from large uncertainties. For the QCD $\bar \theta$ term, however, reasonably precise values are known, because the  $\bar \theta$ term is related to CP-conserving quark masses through a chiral rotation \cite{Crewther:1979pi,Mereghetti:2010tp}. This means that the non-perturbative QCD matrix elements that connect, for example, $\bar g_0$ to $\bar \theta$, are related to non-perturbative QCD matrix elements that connect meson and baryon masses to quark masses. Since the latter are known, from measurements or from lattice QCD, we control the $\bar \theta$ term relatively well. 

We consider the $\bar \theta$ term through a complex quark mass term
\begin{equation}
\mathcal L_\theta = \bar \theta\,m_*\,\bar q i \gamma^5 q\,,
\end{equation}
where we used $\bar \theta \ll 1$. Here $q=(u\,d\,s)^T$ and we introduced the reduced quark mass
\begin{equation}
m_* = \frac{m_u m_d m_s}{m_s(m_u+m_d) + m_u m_d} = \frac{\bar m(1-\varepsilon^2)}{2+\frac{\bar m}{m_s}(1-\varepsilon^2)}\,,
\end{equation}
where $\bar m = (m_u + m_d)/2$ is the average light quark mass and $\varepsilon = (m_d-m_u)/(m_u+m_d)$ is a dimensionless measure of the quark mass difference. For most purposes it is sufficient to just consider the up and down quarks and $m_*\simeq \bar m(1-\varepsilon^2)/2 \simeq \bar m/2$ since $\varepsilon^2 \ll 1$. We use the determination from the FLAG review \cite{FlavourLatticeAveragingGroupFLAG:2024oxs}
\begin{equation}
\bar m = 3.387[39]\,\mathrm{MeV}\,,\qquad \frac{m_s}{\bar m} = 27.42[12]\,,\qquad m_s = 92.4[1.0]\,\mathrm{MeV}\,,\qquad \varepsilon = 0.347[17]\,.
\end{equation}

A detailed determination of the LECs in Eq.~\eqref{eq:gbarterms} based on SU(3) $\chi$PT at next-to-next-to-leading order was performed in Ref.~\cite{deVries:2015una}. We use these results here, but update the relevant matrix elements to the most recent lattice QCD results. This leads to
\begin{equation}\label{g0theta}
\bar{g}_0(\bar \theta) = - \frac{\delta m_N}{2 F_\pi}\frac{m_* \bar \theta}{\bar m \varepsilon} = -17.2[2.0]\cdot 10^{-3}\,\bar \theta\,,
\end{equation}
where $\delta m_N = 2.54[21]\,\mathrm{MeV}$ describes the strong part of the neutron-proton mass splitting obtained from lattice QCD \cite{FlavourLatticeAveragingGroupFLAG:2024oxs}. The relation between $\bar g_0$ and $\delta m_N$ holds up to next-to-next-to-leading-order chiral corrections, which are covered by the uncertainty. 

A similarly robust relation holds for the CP-odd nucleon-$\eta$ coupling \cite{deVries:2015una}. Here we have
\begin{equation}
   \bar{g}_{0\eta}(\bar \theta) = \frac{1}{\sqrt{3}F_\eta}\left(\frac{\sigma_{Ns}}{m_s}-\frac{\sigma_{Nl}}{2\bar m}\right)2 m_* \bar \theta =-117[36]\cdot 10^{-3}\,\bar \theta\,,
\end{equation}
where the uncertainty is dominated by next-to-next-to-leading-order (N$^2$LO) chiral corrections. 
$\sigma_{Ns}$ and $\sigma_{Nl}$ are nucleon sigma terms that describe, respectively, the contribution from the strange and light quark masses to the average nucleon mass. We use $\sigma_{Ns} = 44.9[6.4]$ MeV \cite{FlavourLatticeAveragingGroupFLAG:2024oxs} and $\sigma_{Nl}=59.1[3.5]$ MeV \cite{Hoferichter:2015dsa}, as obtained from a Roy-Steiner analysis. Lattice QCD calculations for $\sigma_{Nl}$ are less precise due to excited-state contaminations \cite{Gupta:2021ahb}. 

The $\bar \theta$ term breaks chiral symmetry but conserves isospin symmetry. This implies it can induce $\bar g_0$ and $\bar g_{0\eta}$ directly, but requires an extra insertion of the quark mass difference to induce the isospin-symmetry-breaking interaction $\bar g_1$. This both suppresses this LEC and makes the connection to the hadron spectrum more complicated. While LO relations exist for $\bar g_1$, they are broken by NLO corrections, leading to a larger uncertainty. We use the prediction of Ref.~\cite{Bsaisou:2012rg}, which is based on $SU(2)$ $\chi$PT and a resonance saturation estimate of a short-distance contribution,
\begin{equation}
   \bar{g}_{1}(\bar \theta) = 3.4[1.5]\cdot 10^{-3}\,\bar \theta\,,
\end{equation}
with roughly $50\%$ uncertainty. 

The above values for $\bar g_0$ and $\bar g_{0\eta}$ are in reasonable agreement with those used in Ref.~\cite{PospelovThO}, respectively $-17\cdot 10^{-3}\,\bar \theta$ and $-85 \cdot 10^{-3}\,\bar \theta$, without an uncertainty estimate. Ref.~\cite{PospelovThO} did not consider $\bar g_1$, but, because numerically $|\bar g_1| \sim |\bar g_{0\eta}| (m_\pi^2/m_\eta^2)$, its contribution will turn out to be comparable in size to the $\eta$ contribution.

\section{Two-photon exchange processes contributing to \texorpdfstring{$C_\mathrm{SP}$}{CSP}}\label{sec:calculations}

\subsection{Meson exchange diagrams}\label{subsec:LO}
\begin{figure}[t]
    \centering
    \begin{subfigure}[b]{0.3\textwidth}
         \centering
         \includegraphics[width=\textwidth]{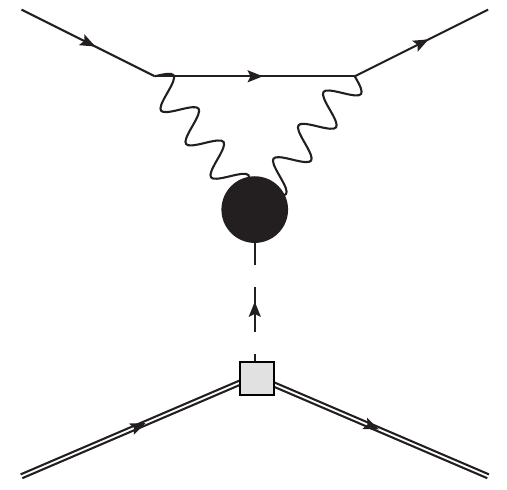}
         \caption{ME loop contribution.}
         \label{fig:LOdiagramforpaperloop}
     \end{subfigure}
     \hspace{0.2\textwidth}
     \begin{subfigure}[b]{0.3\textwidth}
         \centering
         \includegraphics[width=\textwidth]{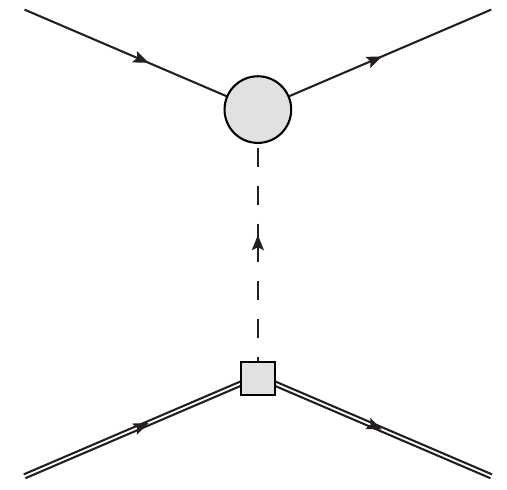}
         \caption{ME counterterm contribution.}
         \label{fig:LOdiagramforpaperCT}
     \end{subfigure}
        \caption{ME diagrams contributing to $C_\mathrm{SP}$. We denote electrons by single and nucleons by double straight lines, pions and eta mesons by dashed lines, and photons by waving lines. The grey square stands for a CP-violating nucleon-nucleon-meson interaction, the black dot for the WZ photon-meson coupling, and the grey dot for the meson-lepton counterterm, which renormalizes the  photon loop.}
    \label{fig:LOdiagramforpaper}
\end{figure}

We now turn to the actual calculation of the diagrams in Fig.~\ref{fig:LOdiagramforpaper}. The CP-odd electron-nucleon amplitude can be written as  
\begin{equation}
    i\mathcal{A}^\text{ME}(q^2) = -\frac{\alpha^2 m_e}{4 \pi^2 F_\pi m_\pi^2} \bar{u}(p_e') \gamma^5 u(p_e) \left[ (\pm\bar{g}_0  + \bar{g}_1) \mathcal{B}^\pi(q^2, m_e) + \frac{F_\pi}{\sqrt{3} F_\eta}\frac{ m_\pi^2}{ m_\eta^2} \bar{g}_{0 \eta} \mathcal{B}^\eta(q^2, m_e) \right]\,,
    \label{eq:LOamplitude}
\end{equation}
where $p_e$ and $p_e'$ are the (on-shell) momenta of the incoming and outgoing electron line, respectively, $q=p_e'-p_e$, and the $\pm$ indicates a $+$ for protons and a $-$ for neutrons. We have not written the heavy-nucleon two-component spinors. The loop function is given by\footnote{Earlier literature exhibits some variation in the constant term in $\mathcal{B}^P$ because of a manifestation of scheme dependence, caused by the definition of the $\gamma^5$ matrix in dimensional regularization \cite{muH,Ramsey-Musolf:2002gmi}. Any constant term can be absorbed into $\chi^P$ and we choose to be consistent with Ref.~\cite{Husek:2015wta, Masjuan:2015cjl,Ametller:1993we}.}
\begin{equation}
    \mathcal{B}^P(q^2, m_e)= 5 +  3 \left(L - \ln{\frac{m_e^2}{\mu^2}}\right) - 2 \chi^P(\mu) - \frac{1}{\beta_e} \left( \text{Li}_2(z_e) - \text{Li}_2\left(\frac{1}{z_e}\right) +i\pi \ln{(-z_e)} \right)\,,
    \label{eq:LOamplitudegistHusek}
\end{equation}
for $P=\{\pi,\eta\}$, $\beta_e = \sqrt{1-4m_e^2/q^2}$, and $z_e = -(1-\beta_e)/(1+\beta_e)$. We have defined the standard combination
\begin{equation}
L =\frac{1}{\varepsilon} -\gamma_E + \ln{4\pi}\,,
\end{equation}
in dimensional regularization, where $\varepsilon = 2-d/2$ and $\gamma_E$ is the Euler-Mascheroni constant. 
Since $q^2<0$, the loop function $\mathcal{B}^P$ is always real. The interactions in Eq.~\eqref{eq:CSPLagrangian} contain no derivatives and correspond to the limit\footnote{Essentially, this implies expanding in $q^2/(4m_e^2)$, which is not necessarily a good approximation for electrons in heavy paramagnetic molecules. Going beyond the $q^2\rightarrow 0$ limit requires the calculation of the long-distance CP-odd electron-nucleus potential and its insertion into molecular many-body computations. Such calculations are currently not available, but this is under investigation.} $q^2\rightarrow 0$, for which
\begin{equation}
    \mathcal{B}^P(0, m_e) = 5 +  3 \left( L - \ln{\frac{m_e^2}{\mu^2}}\right) - 2 \chi^P(\mu) + \dots
\end{equation}

If we set $\mu=m_\rho$, the mass of the $\rho$ meson, and drop the non-logarithmic terms, we obtain the result from Ref.~\cite{PospelovThO}. 

We write the counterterm as $\chi^P(\mu) = (3/2)L + \chi_{\mathrm{fin}}^P(\mu)$ to eliminate the divergence. The finite part depends on the renormalization scale $\mu$ in such a way that the total result is $\mu$-independent. At a chosen $\mu$, $\chi_{\mathrm{fin}}^P(\mu)$ can be obtained by a fit to the measured branching ratios of the rare meson decays $\pi^0 \rightarrow e^+ e^-$ and $\eta^0 \rightarrow \mu^+ \mu^-$. To do so, we consider the ratios
\begin{equation}
    R_{Pll} = \frac{\Gamma(P \rightarrow l^+ l^-)}{\Gamma(P \rightarrow \gamma \gamma)} = \frac{\alpha^2 m_l^2}{2\pi^2 m_P^2} \sqrt{1 - \frac{4m_l^2}{m_P^2}} |\mathcal{B}^P(m_P^2,m_l)|^2\,,
    \label{eq:RP}
\end{equation}
as these are expected to be less sensitive to higher-order chiral corrections than the individual rates \cite{muH}. 

We equate $R_{Pll}$ to the corresponding experimental values at $\mu=m_\rho$, because the two-loop QED radiative corrections (rad) to $P \rightarrow l^+ l^-$ given in Ref.~\cite{Vasko:2011pi} $-$ which are not included in Eq.~\eqref{eq:RP} and should hence be removed from the experimental values too $-$ are computed at this value of $\mu$. We use the recent NA62 result $\mathrm{BR}(\pi^0 \rightarrow e^+ e^-, \text{ no rad}) = 6.22[0.39] \cdot 10^{-8}$ \cite{NA62:2024rxx}, which agrees with the full branching ratio from Ref.~\cite{PDG} when modified using the radiative corrections from Ref.~\cite{Vasko:2011pi}. With this method we obtain\footnote{We take $\chi_\text{fin}^{\eta \mu \mu} \approx \chi_\text{fin}^{\eta ee}$, as expected from $\chi$PT,  because so far only an upper bound on $\mathrm{BR}(\eta \rightarrow e^+ e^-)$ has been measured \cite{PDG}. The corresponding uncertainty is small compared to the $\sim 30\%$ uncertainty on the $\bar{g}_{0\eta}$ discussed in Sec.~\ref{sec:thetaLECs}.} $\mathrm{BR}(\eta \rightarrow \mu^+ \mu^-, \text{ no rad}) = 5.9[0.8] \cdot 10^{-6}$. We use $\mathrm{BR}(\pi^0 \rightarrow \gamma \gamma) = 98.823[0.034] \%$ and $\mathrm{BR}(\eta \rightarrow \gamma \gamma) = 39.36[0.18] \%$ \cite{PDG}. 

As the ratio $R_{Pll}$ depends quadratically on $\chi_\text{fin}^P$, two solutions are obtained:
\begin{align}
\begin{split}
    \chi_\text{fin}^{\pi}(\mu=m_\rho) &= -17.5[1.4] \vee 2.6[1.4]\,,\\
    \chi_\text{fin}^{\eta}(\mu=m_\rho) &= 1.6[1.0] \vee 8.1[1.0]\,.
    \label{eq:mychifitresults}
\end{split}
\end{align}

We will choose between the two solutions for $\chi_\text{fin}^\pi$ and $\chi_\text{fin}^\eta$ by comparing to theoretical predictions. Ref.~\cite{Husek:2015wta} presents theoretical estimates for $\chi_\text{fin}^\pi$ in various underlying resonance models, which all give results between 2 and 3, in agreement with NDA. We therefore choose $\chi_\text{fin}^{\pi} = 2.6[1.4]$. Since $\chi$PT predicts $\chi_\text{fin}^{\pi} = \chi_\text{fin}^{\eta}$, we use $\chi_\text{fin}^{\eta} = 1.6[1.0]$, which agrees with the $\chi$PT prediction within uncertainties. This value is somewhat smaller than obtained in resonance models \cite{Masjuan:2015cjl, Ametller:1993we}, but not too much, considering the large uncertainties. Choosing the other value of $\chi_\text{fin}^\eta$ will only cause a small change in the total value of $C_\mathrm{SP}$. With the fitted values we obtain for the $\mu$-independent loop functions
\begin{align}
\begin{split}
    \mathcal{B}^\pi(0, m_e)&= 43.8[2.7]\,,\\
     \mathcal{B}^\eta(0, m_e)&= 45.8[2.0]\,,
\end{split}
\end{align}
which are completely dominated by the logarithm, with the counterterms contributing only at the $5\%$ level. 

To summarize our discussion of the ME diagrams, we give the expressions for the CP-odd semileptonic couplings,
\begin{align}
\begin{split}
    C_{\mathrm{SP}}^{p,\text{ME}} \frac{G_F}{\sqrt{2}} &= \frac{\alpha^2 m_e}{4 \pi^2 F_\pi m_\pi^2} \left[ \left( \bar{g}_1 + \bar{g}_0 \right) \mathcal{B}^\pi(0,m_e) + \frac{F_\pi}{\sqrt{3} F_\eta}\frac{m_\pi^2}{ m_\eta^2} \bar{g}_{0\eta} \mathcal{B}^\eta(0,m_e) \right]\,, \\
    C_\mathrm{SP}^{n,\text{ME}} \frac{G_F}{\sqrt{2}} &= \frac{\alpha^2 m_e}{4 \pi^2 F_\pi m_\pi^2} \left[ \left( \bar{g}_1 - \bar{g}_0 \right) \mathcal{B}^\pi(0,m_e)  + \frac{F_\pi}{\sqrt{3} F_\eta}\frac{ m_\pi^2}{ m_\eta^2}  \bar{g}_{0\eta} \mathcal{B}^\eta(0,m_e) \right]\,.
    \label{eq:CSPpCSPnME}
\end{split}
\end{align}

\begin{table}[t]
\begin{adjustbox}{tabular=l|ccccc,center}
$C_{\mathrm{SP}}^{\mathrm{ME}}$ & $\bar{g}_0$ & $\bar{g}_{0\eta}$  & $\bar{g}_1$ & Total \\ \hline
Proton  & $-3.77[49]$ & $-0.72[22]$ & $0.74[33]$ & $-3.74[61]$ \\

Neutron   & $\phantom{-}3.77[49]$ & $-0.72[22]$ & $0.74[33]$ & $\phantom{-}3.79[65]$\\
\hline
Nucleus average & $\phantom{-}0.75[10]$  & $-0.72[22]$ & $0.74[33]$ &\phantom{-} $0.78[42]$ 
\end{adjustbox}
\caption{Contributions to $C_\mathrm{SP}$ from  ME diagrams at the nucleon level as well as averaged over the nucleon content of a typical heaviest atom in a diatomic paramagnetic molecule, with $Z/A = 0.4$ and $N/A = 0.6$. We show the contributions from the $\bar{g}_0$, $\bar{g}_{0\eta}$, and $\bar{g}_1$ couplings separately, as well as their sum. All entries should be multiplied by $10^{-2}\, \bar{\theta}$.}
\label{tab:CsLOresults1}
\end{table}

Numerical values of $C^{n,p}_{\mathrm{SP}}$ in terms of $\bar \theta$ are shown in Table~\ref{tab:CsLOresults1}. For a single neutron or proton, the $\bar g_0$ contribution is the largest, with the $\bar g_{0\eta}$ and $\bar g_1$ each contributing at the $20\%$ level. This confirms the expectation discussed in Sec.~\ref{sec:thetaLECs}. 

That being said, the contribution from $\bar g_0$ is isovector and, unlike the $\bar g_{0\eta}$ and $\bar g_1$ contributions, has a relative sign between the proton and neutron. The relevant quantity is
$C_\mathrm{SP}^\text{ME}$ averaged over the nucleon content of the heaviest atom in a diatomic paramagnetic molecule:
\begin{equation}
    C_{\mathrm{SP}}  =  \left( \frac{Z}{A} C_{\mathrm{SP}}^p + \frac{N}{A} C_{\mathrm{SP}}^n \right)\,.
    \label{eq:CSPaveraged}
\end{equation}

As such, the $\bar g_0$ piece is suppressed by $(Z-N)/A \simeq -0.2$ for atoms such as Th, Hf, or Ba, making it comparable to the unsuppressed $\bar g_{0\eta}$ and $\bar g_1$ pieces. Putting everything together, we find that the nucleus-averaged value,
\begin{equation}
C^\text{ME}_{\mathrm{SP}}= 0.78[42]\cdot 10^{-2}\,\bar \theta\,,
\end{equation}
is relatively small and uncertain. Our central value is roughly $8$ times bigger than that of Ref.~\cite{PospelovThO}, who quoted $0.001\,\bar \theta$ with a large but unspecified uncertainty. Our result is significantly larger but considering the large uncertainty not in obvious disagreement. The difference mainly stems from the values of the CP-odd LECs and from the inclusion of the $\bar g_1$ diagrams, while the inclusion of the counterterms in Eq.~\eqref{eq:CT} only plays a marginal role.

\subsection{Pion loop diagrams}\label{subsec:NLO}
\begin{figure}[t]
    \centering
    \begin{subfigure}[b]{0.28\textwidth}
         \centering
         \includegraphics[width=\textwidth]{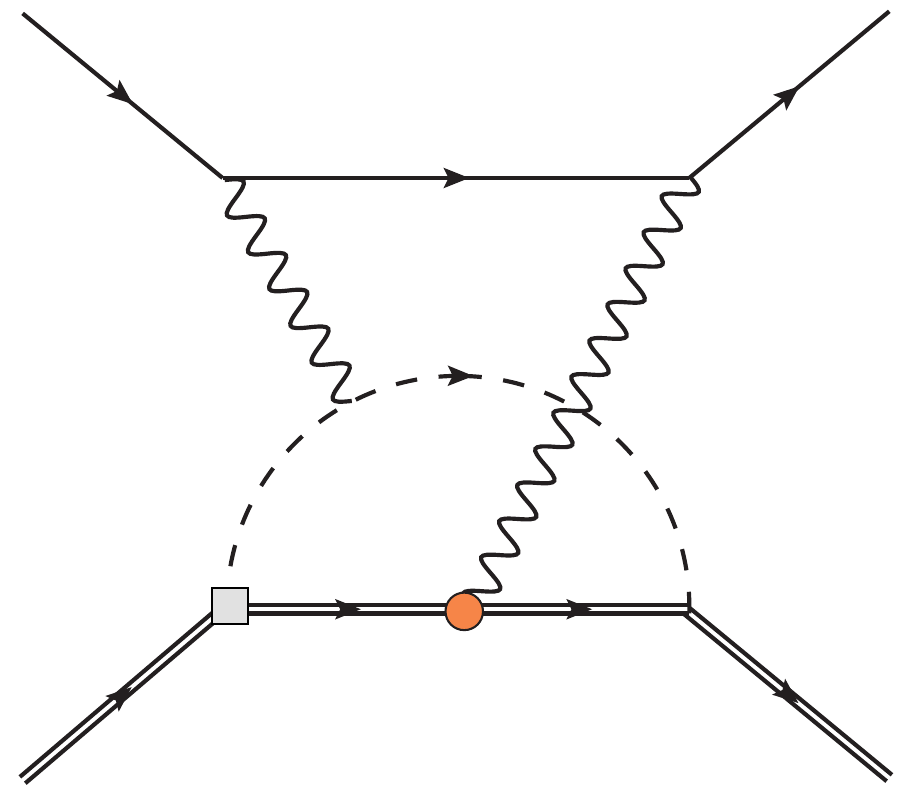}
         \caption{}
         \label{fig:nonzeroNLO}
     \end{subfigure}
     \hfill
     \begin{subfigure}[b]{0.28\textwidth}
         \centering
         \includegraphics[width=\textwidth]{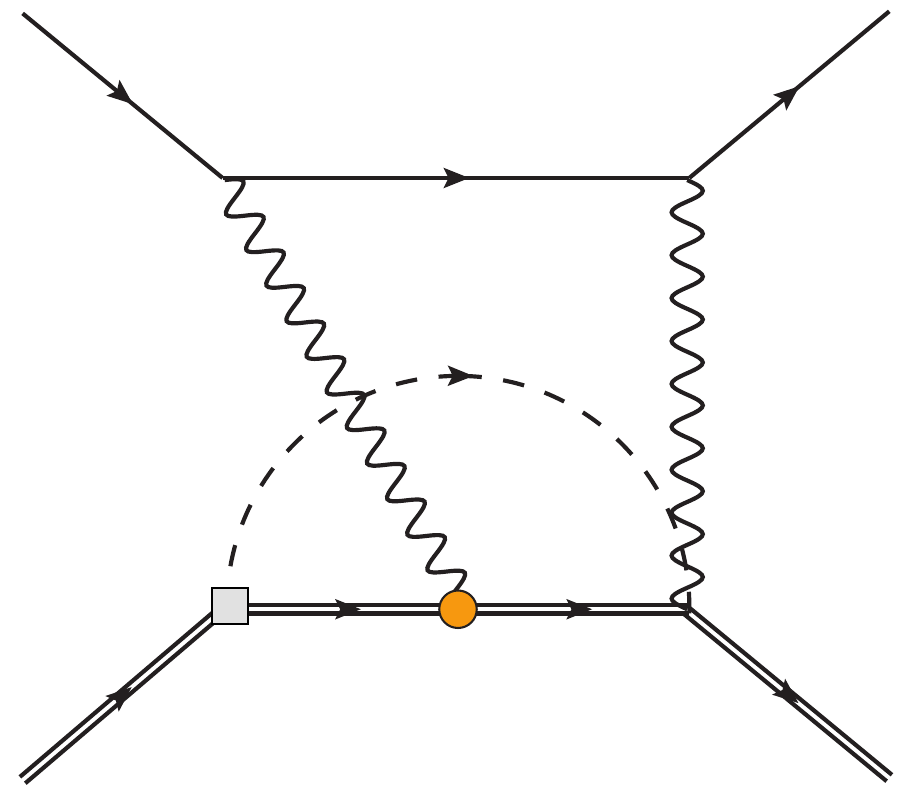}
         \caption{}
         \label{fig:zeroNLO4pt}
     \end{subfigure}
     \hfill
     \begin{subfigure}[b]{0.28\textwidth}
         \centering
         \includegraphics[width=\textwidth]{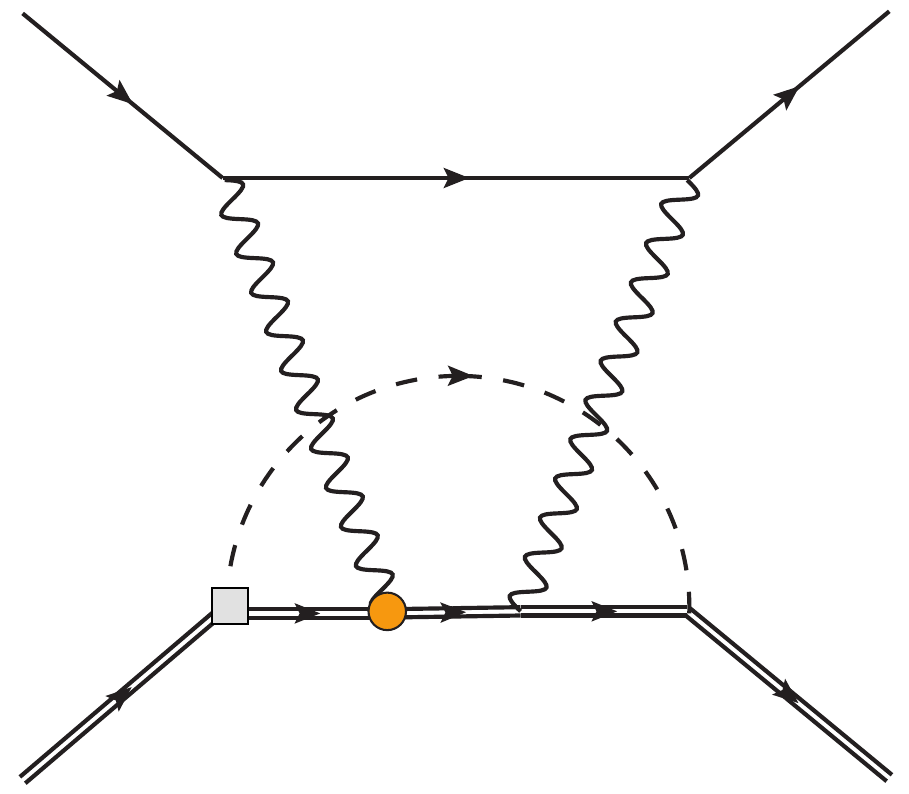}
         \caption{}
         \label{fig:zeroNLOtwotoN}
     \end{subfigure}
        \caption{PL diagrams contributing to $C_\mathrm{SP}$. We denote electrons by single and nucleons by double straight lines, pions and eta mesons by dashed lines, and photons by waving lines. The grey square stands for a CP-violating nucleon-nucleon-pion interaction, and the orange dot for the nucleon magnetic moment. For each type of diagram only one topology is shown.}
    \label{fig:fullNLOdiagrams}
\end{figure}

Considering the accidental suppression of the LO ME diagrams and the resulting large uncertainty, we should consider the PL diagrams that contribute at NLO. We show the relevant PL diagrams contributing to $C_\mathrm{SP}$ in Fig.~\ref{fig:fullNLOdiagrams}. These are all two-loop diagrams that, broadly speaking, can be divided into a photon loop and a pion loop, containing a CP-violating $\bar{g}_0$ coupling as in Eq.~\eqref{eq:gbarterms} and a nucleon magnetic moment vertex as in Eq.~\eqref{eq:anommagnmom}. At this order, the nucleon magnetic moment must enter to ensure that the amplitude is independent of the nucleon spin, as is required to contribute to $C_\mathrm{SP}$. 

The PL diagrams in Fig.~\ref{fig:fullNLOdiagrams} were computed in Ref.~\cite{PospelovThO} by factorizing the bottom (the pion loop) and upper half (the photon loop) of the diagrams. This greatly simplifies the calculation, but leads to a divergent amplitude because the upper half, after integrating out the pions, is the same diagram as computed in Sec.~\ref{subsec:LO}. This problem can be solved by first integrating out pions and matching to a pionless theory, and then computing the renormalization group evolution of the $C_\mathrm{SP}$ interaction to lower energies. This would provide the leading logarithmic corrections, but we are interested in the finite, non-logarithmic, matching contributions as well. We therefore perform a two-loop calculation which provides the complete result.

When considering all possible topologies, we find that the contribution of the diagrams in Fig.~\ref{fig:zeroNLO4pt} and \ref{fig:zeroNLOtwotoN} to $C_\mathrm{SP}$ vanishes. To compute the two-loop diagram in Fig.~\ref{fig:nonzeroNLO}, we have made a number of simplifying steps and assumptions. We briefly outline these here, and give more details in Appendix \ref{app:NLO}. Since $C_\mathrm{SP}$ is a spin-independent coupling, we have only kept contributions with an even number of nucleon spins. We use HB$\chi$PT and neglect the energies of the in- and outgoing nucleons. We only take into account terms up to first order\footnote{One subtlety is that we do keep a term proportional to the electron momentum in the denominator of the electron propagator, because it serves as an infrared regulator.} in the ultrasoft scales $m_e$ and $q = p_e'-p_e$. Finally, we again evaluate the amplitude at $q^2=0$, but this only neglects terms of $\mathcal O(q^2/m_\pi^2)$, which are small even for large molecules. 

For the total two-loop amplitude we then obtain
\begin{align}
\begin{split}
    i \mathcal{A}^\text{PL} = \frac{\alpha^2}{4\pi^2}\frac{m_e \bar{g}_0 g_A}{m_N F_\pi m_\pi} \frac{\mu_p - \mu_n \mp (\mu_p + \mu_n) }{\mu_N} \, \bar{u}(p_e') \gamma^5 u(p_e)\, I(m_e/m_\pi),
\label{eq:intermediatePLampl}
\end{split}
\end{align}
where the $\mp$ indicates a $-$ for protons and a $+$ for neutrons and we express $\mathcal{A}^\text{PL}$ in terms of a loop integral $I$ that depends on the ratio of masses $ m_e / m_\pi$. We were not able to compute $I(r)$ in closed form and instead write it in terms of the integral
\begin{eqnarray}
    I(r) &=& \int_0^1 dx \int_0^\infty d\bar{\lambda} \int_0^1 da \int_0^{1-a} db \int_0^{1-a-b} dc  \nonumber\\
    &&\times \frac{\bar{\lambda}^2}{\left[ (a+b) (\bar{\lambda}^2 + 1) + \bar \lambda^2 \frac{[a+(b-a)x]^2}{x(1-x)} + c^2  x(1-x)\,r^2 \right]^2}\,.
    \label{eq:paramintegral}
\end{eqnarray}

In this expression $x$, $a$, $b$, and $c$ are Feynman parameters, while $\bar{\lambda} = \lambda/m_\pi$ is a dimensionless Schwinger parameter.

\begin{figure}[t]
    \centering
    \includegraphics[width=0.6\linewidth]{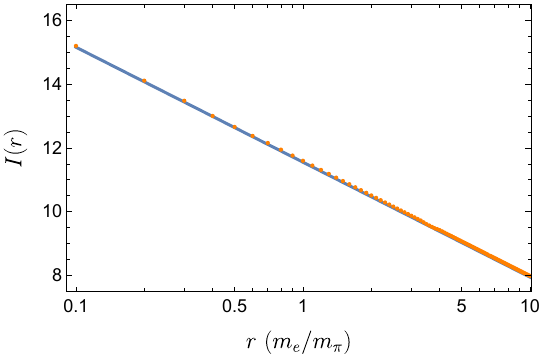}
    \caption{Fit of the parameter integral $I(r)$ over $x$, $\bar{\lambda}$, $a$, $b$ and $c$ in Eq.~\eqref{eq:paramintegral} to a function $C_1 (\log{r}+C_2)$ ($r$ in units of $m_e/m_\pi$). The orange dots show the integral result at various values of $r$, and the blue line shows the best fit result in this range, with $C_1 \simeq -1.57 \simeq - \frac{\pi}{2}$ and $C_2 \simeq -1.77$.}
    \label{fig:rfit}
\end{figure}

As anticipated, $I(r)$ is free of UV or IR divergences and can be computed numerically as $I(m_e/m_\pi)\simeq 11.6$. Factorization of the loops suggests that $I(r)$ has a logarithmic dependence on $r$ \cite{PospelovThO}. Hence, we show a fit of the full two-loop parameter integral $I(r)$ to the functional form $C_1 ( \log  r + C_2)$ in Fig.~\ref{fig:rfit}. For $r \in \{0.1(m_e/m_\pi), 10 (m_e/m_\pi)\}$, we find $C_1 \simeq -1.57 \simeq - \frac{\pi}{2}$ and $C_2 \simeq -1.77$, which provides an excellent fit. We will therefore use 
\begin{equation}
    I(r) = \frac{\pi}{2}\left[\log \left(\frac{1}{r}\right)+1.77 \right]\,,
\end{equation}
in our expressions below.

 We can now write down the contribution from the PL diagrams to $C_{\mathrm{SP}}^{p,n}$:
\begin{align}
\begin{split}
    C_\mathrm{SP}^{p,\text{PL}} \frac{G_F}{\sqrt{2}} &=\phantom{-}  \frac{\mu_n}{\mu_N}\frac{\alpha^2}{4\pi} \frac{m_e g_A \bar{g}_0}{F_\pi m_\pi m_N} \left(\log{\left( \frac{ m_\pi}{m_e}\right)} + 1.77 \right)\,, \\
    C_\mathrm{SP}^{n,\text{PL}} \frac{G_F}{\sqrt{2}} &= -\frac{\mu_p}{\mu_N}\frac{\alpha^2}{4\pi} \frac{m_e g_A \bar{g}_0}{F_\pi m_\pi m_N} \left(\log{\left( \frac{ m_\pi}{m_e}\right)} + 1.77 \right).
    \label{eq:CSPpCSPnPL}
\end{split}
\end{align}

As mentioned, Ref.~\cite{PospelovThO} used factorization to compute the same diagrams. It only kept the leading logarithm and set the renormalization scale, appearing in Eq.~\eqref{eq:LOamplitudegistHusek}, to $\mu=m_\pi$ by hand. Our full two-loop result closely resembles this result, but we have an overall factor $2/3$ and the extra term $+1.77$ in brackets. Numerically, the constant term is a $30\%$ correction to the logarithmic term. 

Compared to the power-counting estimate in Eq.~\eqref{PCNLO}, our result in Eq.~\eqref{eq:CSPpCSPnPL} is enhanced by a factor of $\pi$ and the large logarithm. In addition, the PL contributions to the proton and neutron have the same sign (because $\mu_n$ and $\mu_p$ have opposite sign) and thus, unlike the ME diagrams proportional to $\bar g_0$, do not cancel in heavy atoms. We discuss the numerical implications in the next section.

\section{Combination of results and impact on EDMs}\label{sec:combinedsummary}

We now combine our ME and PL results for $C_\mathrm{SP}$, averaged over the nucleon content of a general nucleus with $Z$ protons and $N$ neutrons:
\begin{align}
\begin{split}
 C_\mathrm{SP}^\text{ME+PL}\frac{G_F}{\sqrt{2}} = \frac{\alpha^2}{4 \pi^2} \frac{ m_e}{ F_\pi m_\pi^2} \bigg\{ &\left[\frac{Z-N}{Z+N} \bar{g}_0 + \bar g_1 \right]\mathcal{B}^\pi(0,m_e) + \frac{F_\pi}{\sqrt{3} F_\eta}\frac{ m^2_\pi}{ m_\eta^2} \bar{g}_{0\eta} \mathcal{B}^\eta(0,m_e) \\ &+ \bar{g}_0\frac{\pi g_A m_\pi }{m_N} \frac{ Z \mu_n-N \mu_p}{(Z+N) \mu_N} \left[\log{\left(\frac{m_\pi}{m_e}\right)}+1.77 \right] \bigg\}\,.
\end{split}
\end{align}

Plugging in numerical values for all parameters (including the counterterm fit results $\chi_\text{fin}^P$) except for $Z$, $N$, and the CP-odd LECs gives
\begin{align}
\begin{split}
    C_\mathrm{SP}^\text{ME+PL} = &\left( \bar{g}_0 \frac{Z-N}{Z+N} + \bar{g}_1 \right) 2.18[14]_\text{ME} + \bar{g}_{0\eta} \,0.0615[27]_\text{ME} \\
    -& \bar{g}_0 \left( \frac{Z}{Z+N} 0.391[24]_\text{PL} + \frac{N}{Z+N} 0.571[36]_\text{PL} \right)\,.
    \label{eq:CSPforuser}
\end{split}
\end{align}

This expression is one of the main results of this paper. It can be used to interpret general paramagnetic atomic and molecular EDM experiments in terms of the hadronic CP-odd meson-nucleon couplings. 
If we specify to the QCD $\bar \theta$ term, we obtain
\begin{align}
\begin{split}
    C_\mathrm{SP}^\text{ME+PL}(\bar \theta) = \left( 0.00026[399] -0.0309[42] \frac{Z}{Z+N} + 0.0475[59] \frac{N}{Z+N} \right) \bar \theta\,.
 \label{eq:CSPforusertheta}
\end{split}
\end{align}

To study the chiral convergence, we consider the couplings to protons and neutrons separately. This gives
\begin{align}
\begin{split}
    C_\mathrm{SP}^{p,\text{ME+PL}}(\bar \theta) &= (-3.77[49]_{\bar{g}_0} - 0.72[22]_{\bar{g}_{0\eta}} + 0.74[33]_{\bar{g}_1} + 0.68[8]_\text{PL}) \cdot 10^{-2} \bar{\theta} \\
    &= -3.07[56] \cdot 10^{-2} \bar{\theta}\,, \\
    C_\mathrm{SP}^{n,\text{ME+PL}}(\bar \theta)&= (3.77[49]_{\bar{g}_0} - 0.72[22]_{\bar{g}_{0\eta}} + 0.74[33]_{\bar{g}_1} + 0.99[11]_\text{PL}) \cdot 10^{-2} \bar{\theta}\\
    &= 4.77[73] \cdot 10^{-2} \bar{\theta}\,.    \label{eq:LONLOsummarynucleon}
\end{split}
\end{align}

In principle, the chiral expansion appears to be converging properly. The LO ME diagrams involving $\bar g_0$ are indeed the largest. The isospin-breaking contributions from $\bar g_1$ provide order $20\%$ corrections, and the same holds for effects from strange quarks through the $\eta$ contribution. The NLO PL loops provide $-18\%$ and $+26\%$ corrections for protons and neutrons, respectively. The total result at this order has roughly $20\%$ total uncertainty for the individual couplings to protons and neutrons. 

However, this convergence pattern is no longer clear once we average over a typical heavy nucleus. For exact isospin symmetry $N$=$Z$, the LO ME diagrams proportional to $\bar g_0$ would vanish exactly. For atoms and molecules of experimental interest we have $N \simeq 3Z/2 $, such that we do not have an exact cancellation but still a strong suppression. For the nucleus-averaged interaction we obtain\footnote{The errors from the various contributions do not quite add quadratically due to correlations stemming from common parameters.}
\begin{align}
\begin{split}
    C_\mathrm{SP}^\text{ME+PL}(\bar \theta) &= (0.75[10]_{\bar{g}_0} - 0.72[22]_{\bar{g}_{0\eta}} + 0.74[33]_{\bar{g}_1} + 0.86[10]_\text{PL}) \cdot 10^{-2} \bar{\theta} \\
    &= (0.78[42]_\text{ME} + 0.86[10]_\text{PL}) \cdot 10^{-2} \bar{\theta} = 1.63[45] \cdot 10^{-2} \bar{\theta}\,.   \label{eq:LONLOsummaryaverage}
\end{split}
\end{align}
Due to the partial cancellation of the LO term and another partial cancellation between the remainder and subleading ME contributions, we see that the formally NLO PL diagrams actually provide the largest contribution. However, as can be seen from Eq.~\eqref{eq:LONLOsummarynucleon}, this is not due to a failure of the chiral expansion but simply to the isovector nature of the LO contribution. We do not expect that N${}^2$LO contributions would be anomalously big, and we expect such contributions to be captured by the sizeable uncertainty of our final answer. Examples of such higher-order contributions could originate from higher-order corrections to the $\bar{g}_{0,1}$ coupling constants, which Ref.~\cite{deVries:2012ab} found to be of the order of a few percent. The same holds for the two-loop QED radiative corrections to the $P \rightarrow l^+ l^-$ decay relevant to our ME diagrams \cite{Vasko:2011pi}.

To facilitate the comparison of our final value for $C_\text{SP}(\bar\theta)$ to the results of EDM experiments with paramagnetic molecules, which are often interpreted as bounds on $d_e$, we also show the result in Eq.~\eqref{eq:LONLOsummaryaverage} in the form $d_e^\text{equiv} \equiv r C_\text{SP}$ \cite{Pospelov:2013sca}. For HfF$^+$, using the value $r=9.17[52] \cdot 10^{-21}$ e cm \cite{Fleig:2018bsf}, we obtain

\begin{equation}
    d_e^\text{equiv}(\bar\theta) = 1.50[42] \cdot 10^{-22} \,\bar{\theta} 
\text{ e cm}\,.
\end{equation}

Eqs.~\eqref{eq:CSPforusertheta} and \eqref{eq:LONLOsummaryaverage} provide the second main result of this work. They can be used to constrain $\bar \theta$ from paramagnetic EDM experiments with a well-defined theoretical uncertainty. Ref.~\cite{Roussy:2022cmp} cites the following experimental result for $C_\mathrm{SP}$, based on experiments done on HfF$^+$ at JILA:
\begin{equation}
    C_\mathrm{SP} = -1.4[2.2]_\text{stat}[0.7]_\text{syst} \cdot 10^{-10}\,,
    \label{eq:exptCspbound}
\end{equation}
 where, as is appropriate for the QCD $\bar \theta$ term \cite{Choi:1990cn}, it is assumed that $d_e=0$. Combining this with our final result\footnote{For Hf, the values $Z/A = 0.4$ and $N/A = 0.6$ used in Eq.~\eqref{eq:LONLOsummaryaverage} hold up to our precision.} in Eq.~\eqref{eq:LONLOsummaryaverage}, we obtain
\begin{equation}
    \bar{\theta}_{\text{HfF}^+}  = -0.9[1.4]\cdot 10^{-8}\,,
\end{equation}
 or
\begin{equation}
    |\bar{\theta}|_{\text{HfF}^+} < 1.5 \cdot 10^{-8}\,,
    \label{eq:thetabound}
\end{equation}
 with $90\%$ confidence.

The neutron EDM is measured as $d_n = 0.0[1.1]_\text{stat}[0.2]_\text{syst} \cdot 10^{-26}$ e cm \cite{Abel:2020pzs}, and using the lattice QCD value $d_n (\bar{\theta})=-1.48[0.34]\cdot 10^{-3}\,\bar \theta$ e fm \cite{Liang:2023jfj} gives $|\bar \theta| < 1.2 \cdot 10^{-10}$ ($90\%$ confidence). This shows that if  paramagnetic EDM experiments improve by roughly two orders of magnitude, they will become competitive in constraining purely hadronic CP-violating sources like the QCD $\bar \theta$ term.

\section{Conclusions and outlook}\label{sec:conclusions}
We have calculated the CP-violating semileptonic nucleon-spin-independent coupling $C_\mathrm{SP}$ in terms of CP-odd meson-nucleon interactions. The latter are directly induced by hadronic sources of CP violation at the quark-gluon level, such as the QCD $\bar \theta$ term or higher-dimensional operators like quark chromo-EDMs or four-quark operators. 
The actual computation involves two classes of diagrams. Both classes involve the exchange of two photons between the electron and nucleon line, mediated either by an additional meson exchange (ME) or pion loop (PL). For the ME diagrams, we have renormalized the divergent photon loop through a local meson-electron counterterm that we  fitted to measured branching ratios of the rare decays $\pi^0 \rightarrow e^+ e^-$ and $\eta \rightarrow \mu^+ \mu^-$. We have included contributions from three separate CP-odd meson-nucleon couplings: the isospin-conserving $\bar{g}_0$ and $\bar{g}_{0\eta}$, and the isospin-violating $\bar{g}_1$. The contribution from $\bar g_0$ is reduced in heavy atoms because of a $(Z-N)/(Z+N)$ suppression.

In the case of the QCD $\bar \theta$ term, we find strong cancellations between the various ME contributions. The inclusion of $\bar g_1$, although formally subleading, therefore becomes relevant. Our efforts to include the counterterms have proven less fruitful: they only slightly change the result. When the dust settles,  we obtain a value for the averaged $C_\mathrm{SP}^\text{ME}$ which is about eight times larger than that of Ref.~\cite{PospelovThO}, but the associated uncertainties are large.

Proceeding to the PL diagrams, we have carried out a full two-loop calculation. The PL diagrams are subleading compared to ME diagrams for individual neutrons and protons, as expected from the $\chi$PT power counting. For the $\bar \theta$ term, however, they become dominant in systems of experimental interest because of the cancellation of the leading $\bar g_0$ terms. 

Our main results are given in Eq.~\eqref{eq:CSPforuser}, \eqref{eq:CSPforusertheta}, and \eqref{eq:LONLOsummaryaverage}, which can be used to connect paramagnetic EDM measurements to hadronic sources of CP violation. For the $\bar \theta$ term, we used relatively precise values of the CP-odd LECs \cite{deVries:2015una,FlavourLatticeAveragingGroupFLAG:2024oxs} to directly express paramagnetic EDMs in terms of $\bar \theta$. From the currently most stringent paramagnetic EDM experiments, which use HfF$^+$, we obtain a constraint $|\bar{\theta}| < 1.5 \cdot 10^{-8}$ at $90\%$ confidence, roughly two orders of magnitude less stringent than the limit from the neutron EDM experiment. This demonstrates that future paramagnetic molecular EDM experiments have the potential to become the best probe of strong CP violation.

Our results can be used for other sources of hadronic CP violation as well. For example, in case of quark chromo-EDMs we would expect $\bar g_1 \sim \bar g_0$ \cite{Pospelov:2001ys}. For CP-odd chiral-breaking four-quark operators, we even expect $\bar g_1 \gg \bar g_0$ \cite{Dekens:2014jka}. In such cases, the main contribution arises from the ME diagrams proportional to $\bar g_1$, that are not suppressed by $(Z-N)/(Z+N)$. This immediately implies that the ratio of paramagnetic over diamagnetic EDMs could actually be used to disentangle different hadronic CP-violating sources.

Our work should be continued in several directions:
\begin{itemize}
\item Perhaps most important is another contribution to $C_{\mathrm{SP}}$ proportional to the combinations of a nucleon EDM and the nucleon magnetic moment. This box contribution was found to be comparable in size to the PL contribution Ref.~\cite{PospelovThO,Flambaum:2020gou}, but the involved nuclear uncertainty is much larger than for the ME and PL diagrams considered here. We are currently investigating whether an effective field theory approach, also used for radiative corrections to superallowed beta decays \cite{Cirigliano:2024msg} and neutrinoless double beta decay \cite{Dekens:2023iyc}, can be applied to better organize this computation. 
\item It would be prudent to investigate whether we should keep the $q^2/m_e^2$ corrections that enter the ME diagrams. This requires more complicated atomic and molecular structure calculations, involving a longer-range CP-odd electron-nucleus potential.  This work is being initiated.
\item We have mainly focused on the QCD $\bar \theta$ term, because for this source the sizes of CP-odd meson-nucleon LECs are relatively well understood. In recent years, there has been a lot of effort to analyze EDMs in terms of the Standard Model Effective Field Theory \cite{deVries:2012ab,Dekens:2013zca,Kley:2021yhn,Kumar:2024yuu}, but the contributions from hadronic CP-odd quark-gluon operators to paramagnetic EDMs has not been considered. It would be interesting to remedy this based on the expressions derived in this work. 
\end{itemize}

In conclusion, the spectacular progress of the last decade in paramagnetic EDM experiments using polar molecules shows no signs of slowing down. They are becoming so precise that the traditional separation into paramagnetic and diamagnetic EDMs is becoming obsolete. The expressions derived in this work can be used to reinterpret paramagnetic EDM experiments in terms of the QCD $\bar \theta$ term and other hadronic sources of CP violation.

\section*{Acknowledgements}
We thank Robert Berger, Ignacio Agustín Aucar, Lukáš Pašteka, Wouter Dekens and Robin van Bijleveld for helpful discussions. This work was partly funded by the Netherlands Research Council (NWO) under programme XL21.074.

\printbibliography
\newpage
\begin{appendices}

\section{Details of the two-loop pion diagrams}\label{app:NLO}

\begin{figure}[h]
    \centering
    \begin{subfigure}[b]{0.45\textwidth}
         \centering
         \includegraphics[width=\textwidth]{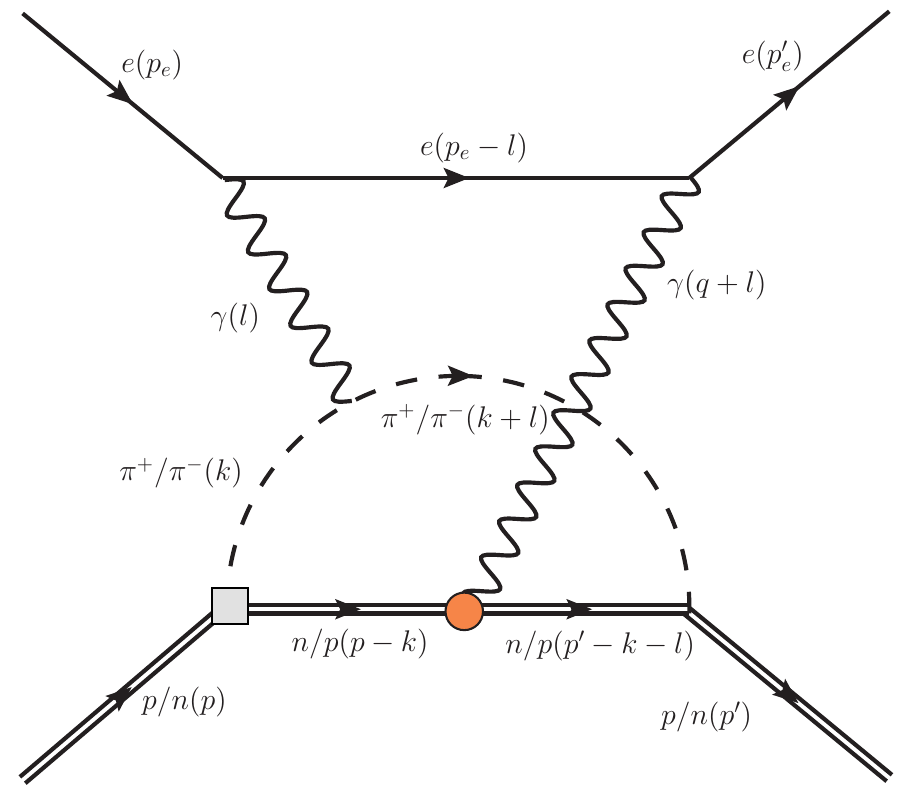}
         \caption{Diagram A.}
         \label{fig:NLOdiagramA}
     \end{subfigure}
     \hfill
     \begin{subfigure}[b]{0.45\textwidth}
         \centering
         \includegraphics[width=\textwidth]{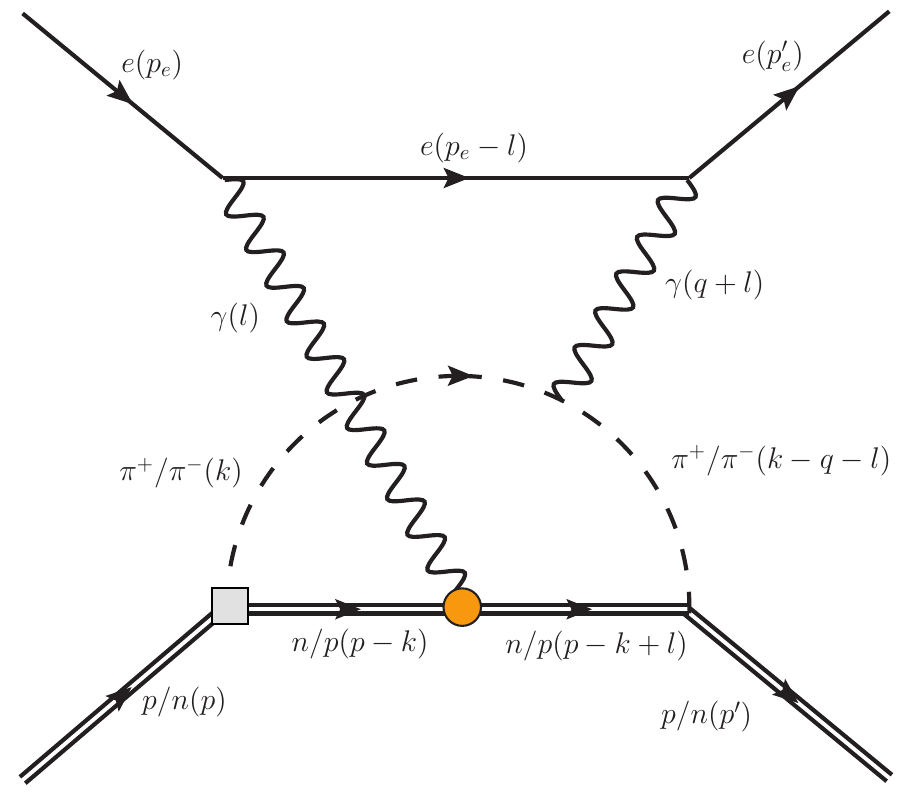}
         \caption{Diagram B.}
         \label{fig:NLOdiagramB}
     \end{subfigure}
        \caption{PL diagrams contributing to $C_\mathrm{SP}$: we show two of the four possible topologies for the only non-zero diagram out of the three classes in Fig.~\ref{fig:fullNLOdiagrams}. The final two topologies, diagrams C and D, can be obtained by switching the CP-even and CP-odd pion-nucleon vertices in diagrams A and B, respectively. We define $q = p_e' - p_e = p - p'$.}
    \label{fig:NLOdiagramswithlabels}
\end{figure}

Here, we provide more details on the full two-loop calculation leading to $\mathcal{A}^\text{PL}$ in Eq.~\eqref{eq:intermediatePLampl} and \eqref{eq:paramintegral}. We use momentum definitions as in Fig.~\ref{fig:NLOdiagramswithlabels}, which shows two of the four possible topologies for the diagram class from Fig.~\ref{fig:nonzeroNLO}, i.e.~the only PL diagram class we have found to give a non-zero contribution to $C_\mathrm{SP}$. From diagrams A and B in Fig.~\ref{fig:NLOdiagramA} and \ref{fig:NLOdiagramB}, respectively, the final two topologies C and D can be obtained by switching the CP-even and CP-odd $\pi NN$ couplings. We show here the calculations for the sum of diagrams A and C. This sum turns out to be equal to that of diagrams B and D. 

As mentioned in the text, only terms with an even number of spins contribute to the spin-independent coupling $C_\mathrm{SP}$. To make sure we only take those terms into account, we rewrite
\begin{equation}
    -S \cdot (k+l) S_\alpha + S_\alpha S \cdot k = -i (k^\beta + \frac{1}{2} l^\beta)\varepsilon_{\alpha\beta\rho\sigma} v^\rho S^\sigma - \frac{1}{4} l^\beta (v_\alpha v_\beta - g_{\alpha \beta})\,,
    \label{eq:rewritespins}
\end{equation}
on the nucleon line. Here $S$ denotes the nucleon spin and $v$ the nucleon velocity. Since the first term in the resulting expression in Eq.~\eqref{eq:rewritespins} contains one nucleon spin, it does not contribute to $C_\mathrm{SP}$ and we drop it. We neglect the terms $v \cdot p$ and $v \cdot p'$ in the nucleon propagators. We are then left with
\begin{align}
\begin{split}
    i \mathcal{A}^\text{PL}(A+C) &= \frac{-e^4 \bar{g}_0 g_A}{8 m_N F_\pi} \epsilon^{3ab} \tau^b (1+\kappa_0+(1+\kappa_1)\tau^3) \tau^a \varepsilon^{\bar{\alpha} \bar{\beta} \alpha \mu} v_{\bar{\alpha}} q_\mu \cdot \\
    &\int \frac{d^d k d^d l}{(2 \pi)^{2d}} l_{\bar{\beta}}(2k+l)_\beta \frac{1}{-v \cdot k} \frac{1}{-v \cdot k - v \cdot l} \frac{1}{l^2} \frac{1}{(q+l)^2} \cdot \\
    &\frac{1}{k^2-m_\pi^2} \frac{1}{(k+l)^2-m_\pi^2}
    \bar{u}(p_e') \gamma_\alpha \frac{(\slashed{p_e}-\slashed{l}+m_e)}{(p_e-l)^2-m_e^2}\gamma^\beta u(p_e)\,,
    \label{eq:amplACJune11}
\end{split}
\end{align}
 where the Pauli matrices act in isospin space and we have not written the heavy-nucleon doublets.

In order to evaluate the $k$ integral
\begin{equation}
    I_{k,\beta}^{AC} = \int \frac{d^d k}{(2\pi)^d} \frac{(2k+l)_\beta}{(-v \cdot k) (-v \cdot k - v \cdot l) (k^2-m_\pi^2)((k+l)^2-m_\pi^2)}\,,
    \label{eq:kintegralAC}
\end{equation}
 we define $\omega = v \cdot l$ and rewrite the product of nucleon propagators through
\begin{equation}
    \frac{1}{(-v \cdot k) (-v \cdot k - \omega)} = \frac{1}{\omega} \left( \frac{1}{-v \cdot k - \omega} - \frac{1}{-v \cdot k} \right)\,.
    \label{eq:rewritenucleonpropagators}
\end{equation}

Then, using Feynman and Schwinger parametrizations as well as symmetry considerations gives
\begin{equation}
    I_{k,\beta}^{AC} = \frac{4}{\omega} \int \frac{d^d k}{(2\pi)^d} \int_0^1 dx \int_0^\infty d\lambda (l(1-2x)+2\lambda v)_\beta \left( \frac{1}{(k^2-\Delta_1)^3} - \frac{1}{(k^2-\Delta_2)^3}\right),
    \label{eq:kintegralAC2}
\end{equation}

\noindent where we have defined $\Delta = m_\pi^2 -x(1-x)l^2+\lambda^2$, $\Delta_1 = \Delta + 2\lambda \omega (1-x)$ and $\Delta_2 = \Delta - 2 \lambda \omega x$. Using standard techniques for the remaining $k$ integral and substituting the result in $\mathcal{A}^\text{PL}(A+C)$ as in Eq.~\eqref{eq:amplACJune11} gives
\begin{align}
\begin{split}
    i \mathcal{A}^\text{PL}(A+C) &= \frac{i e^4 \bar{g}_0 g_A}{4 m_N F_\pi} \epsilon^{3ab} \tau^b (1+\kappa_0+(1+\kappa_1)\tau^3) \tau^a \varepsilon^{\bar{\alpha} \bar{\beta} \alpha \mu} v_{\bar{\alpha}} q_\mu \cdot \\
    &\int \frac{d^d l}{(2 \pi)^{d}} \int_0^1 dx \int_0^\infty d\lambda \frac{1}{\omega} l_{\bar{\beta}} (l(1-2x)+2\lambda v)_\beta  \frac{1}{l^2} \frac{1}{(q+l)^2} \cdot \\
    &\bar{u}(p_e') \gamma_\alpha \frac{(\slashed{p_e}-\slashed{l}+m_e)}{(p_e-l)^2-m_e^2}\gamma^\beta u(p_e) \frac{\Gamma(3-d/2)}{(4\pi)^{d/2}} \left[ \frac{1}{\Delta_1^{3-d/2}} - \frac{1}{\Delta_2^{3-d/2}} \right].
    \label{eq:amplAC}
\end{split}
\end{align}

We now approximate to first order in the momentum transfer $q$, the in- and outgoing electron momenta $p_e$ and $p_e'$, and $m_e$. In the electron denominator $(p_e-l)^2-m_e^2 = l^2 -2 pe\cdot l$, we keep the $2p_e \cdot l$ term, as it will serve as an IR regulator. This then gives
\begin{align}
\begin{split}
    i \mathcal{A}^\text{PL}(A+C) &= \frac{i e^4 \bar{g}_0 g_A}{4 m_N F_\pi} \epsilon^{3ab} \tau^b (1+\kappa_0+(1+\kappa_1)\tau^3) \tau^a \varepsilon^{\bar{\alpha} \bar{\beta} \alpha \mu} v_{\bar{\alpha}} q_\mu \cdot \\
    &\int \frac{d^d l}{(2 \pi)^{d}} \int_0^1 dx \int_0^\infty d\lambda \frac{1}{\omega} l_{\bar{\beta}} (l(1-2x)+2\lambda v)_\beta  \frac{1}{l^4} \cdot \\
    &\bar{u}(p_e') \gamma_\alpha \frac{(-\slashed{l})}{l^2-2p_e \cdot l}\gamma^\beta u(p_e) \frac{\Gamma(3-d/2)}{(4\pi)^{d/2}} \left[ \frac{1}{\Delta_1^{3-d/2}} - \frac{1}{\Delta_2^{3-d/2}} \right].
    \label{eq:amplAC1stO}
\end{split}
\end{align}

Terms involving three $l$'s become 0 by symmetry. The $l_\alpha l_\beta$ terms must be proportional to $g_{\alpha \beta}$, which leaves us with $\varepsilon^{\bar{\alpha} \bar{\beta} \alpha \mu} v_{\bar{\alpha}} v_\beta q_\mu \bar{u}(p_e') \gamma_\alpha\gamma_{\bar{\beta}}\gamma^\beta u(p_e)$. Evaluating this gives $4i m_e \bar{u}(p_e') \gamma^5 u(p_e) - 2iv \cdot q \bar{u}(p_e') \slashed{v} \gamma^5 u(p_e)$. The second term does not lead to a $C_\text{SP}$ structure and can be neglected for $q^2\ll m_e^2$. 

So far, we have been working in $d$ dimensions. However, the remaining integral is finite and we can safely set $d\rightarrow 4$.  This means
\begin{align}
\begin{split}
    i \mathcal{A}^\text{PL}(A+C) &= \frac{- i e^4 m_e \bar{g}_0 g_A}{8 \pi^2 m_N F_\pi \mu_N} (\mu_p - \mu_n - (\mu_p + \mu_n) \tau^3)  \bar{u}(p_e') \gamma^5 u(p_e) \cdot \\
    &\int_0^1 dx \int_0^\infty d\lambda \lambda^2 \int \frac{d^4 l}{(2 \pi)^{4}} \frac{1}{l^2 (l^2-2p_e \cdot l) \Delta_1 \Delta_2}\,,
    \label{eq:amplACisospin}
\end{split}
\end{align}
 where we have also used the definitions of $\mu_{p,n}$ in terms of $\kappa_{0,1}$. Recalling the definitions of $\Delta_1$ and $\Delta_2$, we combine the denominator using standard Feynman tricks, and perform the $d^4l$ integral
\begin{align}
\begin{split}
    i \mathcal{A}^\text{PL} &= \frac{\alpha^2 m_e \bar{g}_0 g_A}{4 \pi^2 m_N F_\pi m_\pi \mu_N} (\mu_p - \mu_n - (\mu_p + \mu_n) \tau^3)  \bar{u}(p_e') \gamma^5 u(p_e)\,\int_0^1 dx \int_0^\infty d\bar{\lambda}  \\
    &\int_0^1 da \int_0^{1-a} db \int_0^{1-a-b} dc \frac{\bar{\lambda}^2}{x^2 (1-x)^2} \frac{1}{\left[ (a+b) \frac{\bar{\lambda}^2 + 1}{x(1-x)} + \bar{\lambda}^2 A^2 + c^2 r^2 + 2\bar{\lambda} c A \frac{E_e}{m_\pi} \right]^2}\,,\label{eq:fullintermediateNLOampl}
\end{split}
\end{align}
where $E_e = v\cdot p_e$. We have included all four possible diagram topologies here, leading to an extra factor 2 (see Fig.~\ref{fig:NLOdiagramswithlabels}). We employed definitions $A = a/x + b/(1-x)$ as well as $\bar{\lambda} = \lambda / m_\pi$ and $r = m_e / m_\pi$. Comparing to Eq.~\eqref{eq:intermediatePLampl} and \eqref{eq:paramintegral} in the text, the only difference with Eq.~\eqref{eq:fullintermediateNLOampl} is the term $2\bar{\lambda}cA\frac{E_e}{m_\pi}$. Considering that $m_e \sim E_e \ll m_\pi$, one would initially expect the terms $c^2 r^2$ and $2\bar{\lambda}cA\frac{E_e}{m_\pi}$ to be negligible in most of the integration parameter space. However, the term $c^2r^2$ can act as an IR regulator in the limit where $a,b,\bar{\lambda} \rightarrow 0$. The term $\sim E_e$ does not serve such a purpose and, evaluating the integral using (partially) numerical methods, we find that the term $2\bar{\lambda}cA\frac{E_e}{m_\pi}$ is indeed negligible and we therefore have not included it in Eq.~\eqref{eq:paramintegral}. 

\end{appendices}

\end{document}